# Kaemika app
# Integrating protocols and chemical simulation

Luca Cardelli

University of Oxford, UK.
luca.a.cardelli@gmail.com

**Abstract.** Kaemika[1] is an app available on the four major app stores. It provides deterministic and stochastic simulation, supporting natural chemical notation enhanced with recursive and conditional generation of chemical reaction networks. It has a liquid-handling protocol sublanguage compiled to a virtual digital microfluidic device. Chemical and microfluidic simulations can be interleaved for full experimental-cycle modeling. A novel and unambiguous representation of directed multigraphs is used to lay out chemical reaction networks in graphical form.

## 1 Introduction

Kaemika is a chemical reaction simulator, including a modern graphical user interface and a functional programming language for platform independent (command line free) operation. It provides basic chemical and stochastic simulation functionality, supporting natural chemical notation and enhancing it with the recursive and conditional generation of chemical reaction networks. It innovates primarily in the integration of liquid-handling protocols with chemical kinetics, providing a unified semantics for laboratory procedures and the evolution of multiple chemicals samples. Based on a previously presented protocol language [1], the app demonstrates its potential by compiling its geometry-free descriptions to a virtual digital microfluidic device that interleaves droplet routing simulation with chemical simulation, for full experimental-cycle modeling. Another contribution is a regular and compact representation of directed multigraphs, which includes a new representation of Petri nets but is further specialized for presenting chemical reaction networks in graphical form.

## 2 Simulation of chemical reaction networks

Kaemika offers deterministic and stochastic simulation of chemical reaction networks, aiming for uniformity of techniques over all expressible reaction networks. Mass action kinetics is used by default, but Hill, Arrhenius, and other kinetics can be expressed via common algebraic and elementary transcendental functions. This includes supplying continuous and discontinuous input waveforms.

---

[1] /'kimika/, a homophone of the Italian word for chemistry

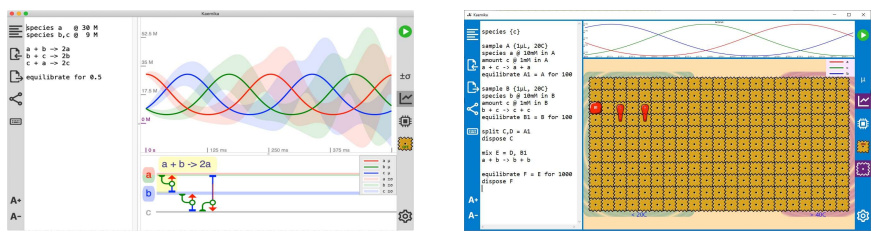

Fig.1: Graphical user interface (macOS left, Windows UWP right). Script editor (left), plot (top), reaction score (Mac, bottom right), microfluidics (Windows, bottom right). Menus and buttons on the sides. Press Play to simulate.

Stochasticity is supported via the Linear Noise Approximation (LNA [8]). Numerical LNA simulations produce predefined displays for standard deviation, variance, coefficient of variation, and Fano factor, and also programmable displays of (variance, etc., of) linear combinations of species using the correct statistics. LNA numerical simulations can be applied to all expressible kinetics, which would be hard to do with other stochastic techniques. LNA is supported also symbolically, providing formal derivatives for the covariance of any pair of species for all expressible kinetics (as long as the kinetic functions are differentiable), which can then be externally studied analytically.

The focus on the LNA technique is due to its general and uniform applicability, and to its relative speed and single-shot operation. The LNA is an approximation of the chemical master equation, and we should complement it with other techniques whenever possible. But the great convenience of the LNA makes it, in my opinion, the default every-day solution, especially in the context of dealing with any (multimolecular, Hill, etc.) reactions that a user may write.

While these simulation techniques are not particularly novel, they are applied in a uniform and consistent way to facilitate experimentation, so that if a user can write down a chemical model, then the tool can in fact simulate it at the click of a button (within the numerical bounds of an ODE solver). An example is the extension of the LNA semantics to liquid-handling simulations (which is novel), where the noise present in a compartment is correctly propagated when the compartment is split or merged with other compartments.

All this functionality is packaged in the interface as a single "play" button for simulation, plus a toggle for the LNA, and a corresponding "stop" button.

## 3  Programmatic generation of networks and protocols

Despite their transparency and simplicity, chemical reaction networks become awkward when they contain many reactions, many repeated subsystems, and many parameters. This is a classical abstraction problem that has been identified and addressed long ago [10] and more recently [12]. Kaemika originated from the desire to build "programmable" (arbitrarily parameterizable) reaction networks, e.g. to study their algorithmic capabilities, and from frustration with existing tools that did not seem to quite meet that need. (An option is to use general programming or mathematical languages, at the cost of largely loosing the notational convenience of chemical reactions.) Can there be an *extension* of chemical reaction notation that makes it fully parameterizable?

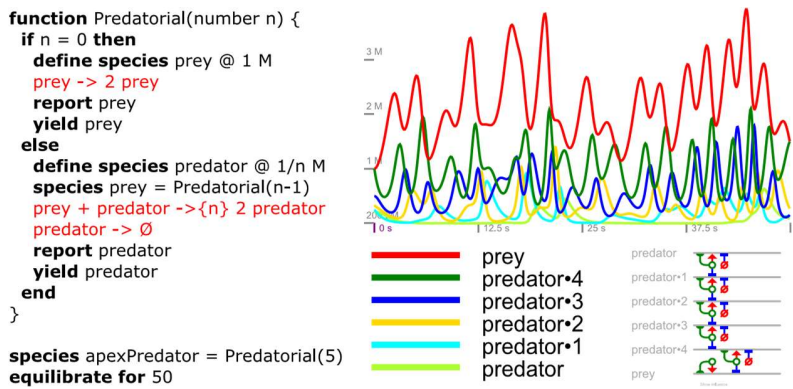

Fig.2: Predator-eat-predator. Left: a program generating a variable-size chemical network, reactions in red. Top: simulation plot for n=5. Bottom center: legend. Bottom right: graphical representation of the generated reaction network.

Kaemika adopts modern concepts from functional programming to solve this problem. First, there is functional programming itself for complete, higher order, abstraction ("Can a species, or a network, be a parameter to a network?"). We then use *nominal* semantics [6] to deal with the generation and lexical binding of an unbounded number of unique chemical species ("If I create new species inside a loop, can I plot them?"). Finally we use an *output monad* [11], which is a somewhat grandiose but systematic scheme for generating a network of chemical reactions from a functional computation ("Can I produce a network whose size is determined by conditional execution?"). All answers are "yes!".

A short example will have to suffice here. The "Predatorial" function in Figure 2 creates a stack of predator-prey relationships in Lotka-Volterra style, and returns the apex predator. To note: (1) the function is recursive; it internally creates new species ('prey', 'predator'), initializes them ('@'), and returns them ('yield'), (2) the new species are 'reported' as they are created, so that they can be plotted, (3) chemical notation (in red) is freely intermixed with flow control, (4) 'equilibrate' runs a simulation and plots it, combining all the reports. The 'equilibrate' statement can be repeatedly invoked. Through some variations of the 'report' statement one can also capture simulation timecourses, recombine them within other simulations, and export them as data.

## 4  Visualization of chemical reaction networks

Automated layout of reaction networks (multigraphs) is usually highly unsatisfactory in the sense of hiding the symmetries of the network, and awkward in the sense of requiring constant panning and zooming. Kaemika uses a new graphical representation of directed multigraphs with multiplicities, which are those needed to unambiguously represent chemical reactions. In first instance, the problem is the same as visually representing Petri nets; even here we appear to be making an original contribution. In addition, catalysts are given a more compact visual representation that extends the basic one for Petri nets.

We call this new representation a *reaction score*. Like a musical score it has a set of horizontal lines, each associated with a chemical species rather than a pitch. Reactions are added to the score in horizontally-bounded vertical tiles. Neither the horizontal nor vertical orders are important (unlike in musical notation), and it is useful to be able to manually or automatically reorder species and reactions to cluster them in different ways. Each reaction $A \to B$ is first recast in the form $C, A^0 \to B^0$ where for each species $s$ if $n*s$ occurs in $A$ and $m*s$ occurs in $B$, then $min(n,m)*s$ are moved into $C$, and the rest are left in $A^0$ or $B^0$ (not both). The reaction $A^0 \to B^0$ is laid out as a Petri net transition and interconnected (the Petri net places are stretched out as horizontal lines, and the transition "bars" are placed vertically, or omitted in 1-input/1-output cases such as in Figure 3). Additional catalytic connections, using a different visual style, are introduced between the species in $C$ and the *stem* (transition) of $A^0 \to B^0$.

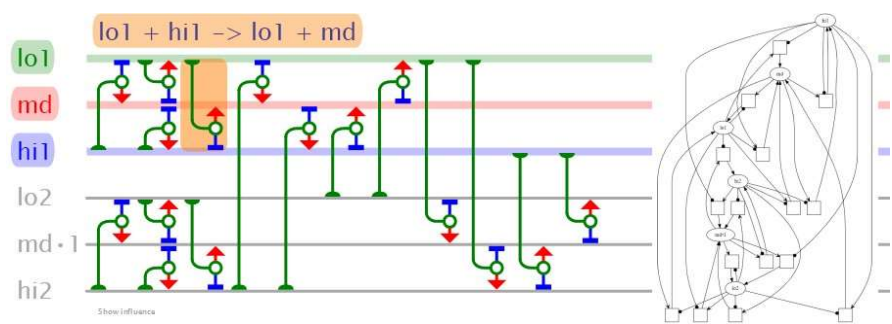

Fig.3: Reaction score. Horizontal lines are species, vertical tiles are reactions. Blue/blunt are reagents, red/sharp are products, green/circle are catalysts. Note some evident substructures and symmetries. On the right, for comparison, is the same multigraph rendered by GraphViz, with ovals for species and squares for reactions, already reduced as in the reaction score via catalytic edges.

This representation is complete (any reaction network can be automatically laid out) and unambiguous (the original reaction network can be recovered from it, except for the reaction rates and initial conditions).

## 5 Protocols and digital microfluidics

The Kaemika system provides a virtual liquid handling device for the simulation and visualization of protocols (Figure 1). We focus on digital microfluidics because of its generality, simplicity, and programmability, in that a single device can execute all the basic liquid-handling protocols [2] [13], and support automated observation of the samples [9].

A Kaemika protocol contains information about the kinetics of the reactions that naturally occur within samples, and also about laboratory manipulations performed on samples [1]. The two are linked because lab operations affect concentrations, volumes, and temperatures, which affect kinetics. Correspondingly, the execution of a Kaemika protocol intertwines the simulation of individual reaction networks with the microfluidic manipulation of the samples, including intertwining the plotting of simulations and the visualization of liquid handling. The state of a sample at the end of a chemical simulation is propagated to the following liquid handling operation, and conversely.

A typical digital microfluidics device has a rectangular array of electrically controlled pads, and some means of adding and removing liquid droplets over its surface. Injection and extraction may by done by hand, or by extruding standard-size droplets from larger on-device reservoirs, or by pumps at the device's periphery. The standard droplet size is around $1\mu L$. Droplets can be moved by changing the electrical properties of adjacent pads, and multiple droplets can be moved in parallel. Droplets can be merged by causing one to move over the pad of another, and split by electrically pulling them in opposite directions. An overhead camera or an on-surface sensing apparatus may provide feedback about the position of the droplets.

In a Kaemika droplet simulation, each "sample" (a container for species and reactions) is represented by a droplet on the device. Mixing, splitting, and disposing of samples is handled by appropriate routing of the droplets over the device surface: this is automatic, and does not require geometric instructions.

Some physical assumptions are needed for timing, for observation, and for the handling of temperatures and volumes. We assume that a region of the device is maintained at a *cool* temperature. All the staging and mixing operation are executed in this region, because chemical reactions are assumed not to be happening during liquid handling: cool temperature and quick execution can approximate those conditions. We also assume that another region of the device is maintained at a *hot* temperature, and an intermediate region is at *warm*, ambient temperature. Times passes, logically, only during "equilibrate" operations, which move droplets into one of the warm or hot regions, according to need, hold them there for the prescribed time, and then move them back to the cool region. Observation capabilities (and subsequent feedback into protocols) are highly hardware dependent [9]: we provide in the language general observability of concentrations, but this will have to be matched to physical device capabilities.

## 6   Implementation and deployment

Kaemika is written in C# using the Visual Studio/Xamarin IDE, and is available on four app stores: Windows UWP, macOS, iOS, and Android. A single Visual Studio solution is used for all platforms, with shared application logic, compiled under either Windows or macOS; the source code is on GitHub [4]. The language syntax is based on the Gold LALR parser generator [5]. The ODE solver is OSLO [7]. The basic simulation functionality, including LNA, is common with many other tools, e.g. [3] [10], which otherwise focus on other modeling aspects. The main Windows and macOS GUI interfaces consist of two similar separate forms; a separate GUI is used for mobile displays, with Xamarin providing a unified interface to Android/iOS. Low-level graphics (lines, splines, fonts, etc.) is shared across Windows/iOS/Android via Skia graphics, but separate from CoreGraphics for macOS. XAML, which subverts lexical scoping, typing, error accountability, and reliability in all applications, is painstakingly circumvented.

In practice, supporting multiple platforms is not hard, and software changes propagate easily across them. Rather, the challenge is navigating the parkourlike registration, provisioning, and app submission procedures of each app store. Still, I strongly advise this path since it has huge benefits for users in terms of tool installation, and also of usability (flawed GUIs are store-rejected). In the end it has huge benefits for developers too, in terms of removing variability of user configurations and all the related distribution and support issues, which I found even more challenging than app store approvals.

## References

1. Abate, A., et al.: Experimental biological protocols with formal semantics. In: CMSB. vol. LNCS 11095, pp. 165–182. Springer (2018)
2. Alistar, M., Gaudenz, U.: Opendrop: An integrated do-it-yourself platform for personal use of biochips. Bioengineering (Basel) **4**(2), 45 (2017)
3. Cardelli, L., Tribastone, M., Tschaikowski, M., Vandin, A.: ERODE: A tool for the evaluation and reduction of ordinary differential equations. In: TACAS (2017)
4. Cardelli, L.: **https://github.com/luca-cardelli/KaemikaXM**
5. Cook, D.: Design and development of a grammar oriented parsing system. MSc Project, California State University Sacramento (2004)
6. Crole, R., Nebel, F.: Nominal lambda calculus: An internal language for fmcartesian closed categories. ENTCS **298**, 93–117 (2013)
7. Dalchau, N.: Open solving library for ODEs **https://www.microsoft.com/enus/research/project/open-solving-library-for-odes/**
8. Ethier, S., Kurtz, T.: Markov processes: characterization and convergence. John Wiley & Sons (2009)
9. Freire, S.: Perspectives on digital microfluidics. Sensors and Actuators A: Physical **250**, 15–28 (2016)
10. Pedersen, M., Phillips, A.: Towards programming languages for genetic engineering of living cells. Journal of the Royal Society Interface **6**, S437–S450 (April 2009)

11. Petricek, T.: What we talk about when we talk about monads. The Art, Science, and Engineering of Programming **2**(2), 12 (2018)
12. Vasic, M., Soloveichik, D., Khurshid, S.: CRN++: Molecular programming language. Natural Computing **19**(1-2) (2020)
13. Willsey, M., et al.: Puddle: A dynamic, error-correcting, full-stack microfluidics platform. In: ASPLOS (2019)